\documentclass[a4paper,12pt,reqno]{article}
\usepackage{amsmath,amsfonts,amssymb,amsthm,dsfont,slashed}
\allowdisplaybreaks

\newcommand{\da}{{\dot \alpha}}
\newcommand{\dbe}{{\dot \beta}}
\newcommand{\dga}{{\dot \gamma}}

\newcommand{\cA}{{\cal A}}
\newcommand{\cB}{{\cal B}}
\newcommand{\cD}{{\cal D}}

\newcommand{\cT}{{\cal T}}
\newcommand{\qd}{{\quad}}

\newcommand{\RS}[2]{\psi_{#2}{}^{#1}}
\newcommand{\bRS}[2]{\5\psi_{#2}{}^{#1}}
\newcommand{\viel}[2]{e_{#2}{}^{#1}}

\newcommand{\bea}{\begin{eqnarray}} \newcommand{\eea}{\end{eqnarray}}
\newcommand{\beann}{\begin{eqnarray*}} \newcommand{\eeann}{\end{eqnarray*}}
\newcommand{\beq}{\begin{equation}} \newcommand{\eeq}{\end{equation}}
\newcommand{\ba}{\begin{array}} \newcommand{\ea}{\end{array}}
\newcommand{\ben}{\begin{enumerate}} \newcommand{\een}{\end{enumerate}}

\newcommand{\om}{\omega}

\newcommand{\4 }{\tilde}
\newcommand{\5}{\bar}
\newcommand{\6}{\partial}
\newcommand{\7 }{\hat}

\begin{document}

\begin{center}
 {\large\bfseries Comment on ``A Candidate for a Supergravity Anomaly''}
 \\[5mm]
 Friedemann Brandt \\[2mm]
 \textit{Institut f\"ur Theoretische Physik, Leibniz Universit\"at Hannover, Appelstra\ss e 2, 30167 Hannover, Germany}
\end{center}

\begin{abstract}
In a recent paper (arXiv:2110.06213v1) it was argued that in 4D supergravity a previously unknown candidate anomaly
exists.  In the present paper it is pointed out that the existence of such a candidate would contradict results available in the literature, and it is substantiated that there is no candidate anomaly of the asserted form.
\end{abstract}

In the recent paper \cite{Dixon:2021ydu} it was argued that in 4D supergravity a previously unknown candidate anomaly 
\bea
\int d^4x \, \epsilon^{\mu\nu\rho\sigma}(c_1\xi_\alpha\6_\mu\RS \alpha\nu\6_\rho A_\sigma+
c_2C\6_\mu\RS \alpha\nu\6_\rho\psi_{\sigma\alpha}+\ldots)
\label{a1}
\eea
exists where $\xi_\alpha$ are supersymmetry ghosts (2-component commuting Weyl spinors), $\RS \alpha\nu$ is the gravitino, $A_\sigma$ is a Yang-Mills gauge field, $C$ is the Yang-Mills ghost corresponding to $A_\sigma$, and $c_1 $ and $c_2$ are constant coefficients (in \cite{Dixon:2021ydu} the notation $V_\sigma$ and $\omega$ is used in place of $A_\sigma$ and $C$). In this context  a candidate anomaly is a local functional with ghost number 1 whose integrand is BRST-invariant up to a total divergence and which is not BRST-exact up to a total divergence in the space of local functions. Hence, \cite{Dixon:2021ydu} asserts that there are coefficients $c_1 $ and $c_2$ and terms (not given in \cite{Dixon:2021ydu}) represented by the ellipses in \eqref{a1} such that the integrand of \eqref{a1} is (i) BRST-invariant up to a total divergence, and (ii) not BRST-exact up to a total divergence. However, this assertion is not proved or properly substantiated in \cite{Dixon:2021ydu}. 

The only argument given in \cite{Dixon:2021ydu} to support the assertion is that ``there is one boundary generator'' (``counterterm'') given by
\bea
\int d^4x \, \epsilon^{\mu\nu\rho\sigma}\psi_{\mu\alpha}\6_\nu\RS \alpha\rho A_\sigma
\label{a11}
\eea
whose BRST-transformation contains, modulo a total divergence, a particular linear combination of the two terms in \eqref{a1} (i.e. a linear combination with particular values of the coefficients $c_1$ and $c_2$), ``but there are two cycles'' in \eqref{a1}, cf. abstract of 
\cite{Dixon:2021ydu}. Hence, \cite{Dixon:2021ydu} asserts that the two terms in \eqref{a1} independently of each other can be completed to a local functional whose integrand is BRST-invariant up to a total divergence (cf. also the text following eq. (18) on page 6 of \cite{Dixon:2021ydu}). This assertion is not proved or substantiated either in \cite{Dixon:2021ydu}.

Summing up, in \cite{Dixon:2021ydu} it is only shown that there is a local functional 
\eqref{a1} (with coefficients $c_1$ and $c_2$ of a particular ratio) whose integrand is BRST-exact up to a total divergence and thus fulfills (i) but not (ii). No evidence is provided that a local functional \eqref{a1} exists which fulfills both (i) and (ii). 

In the following it is pointed out that the existence of a candidate anomaly of the form \eqref{a1} would contradict results derived in \cite{Brandt:1996au}, and that the absence of a candidate anomaly of the form \eqref{a1} can be shown by arguments which do not invoke the more sophisticated results of \cite{Brandt:1996au}. In other words, it will be substantiated that and why there is no candidate anomaly of the form \eqref{a1}.

In \cite{Brandt:1996au} the local BRST cohomology was computed for old \cite{Stelle:1978ye,Ferrara:1978em} and new \cite{Sohnius:1981tp} minimal supergravity, both for pure supergravity and for supergravity coupled to matter multiplets (Yang-Mills multiplets and chiral matter multiplets). The results for candidate anomalies are summarized in section 9 of \cite{Brandt:1996au}. Ten different types of candidate anomalies were found, denoted by type I, II, III, IVa, IVb and Va to Ve, cf. table 9.1 of \cite{Brandt:1996au}. The candidate anomalies of type III, IVb, Vc, Vd and Ve occur only in new minimal supergravity because they are related to the presence of a 2-form gauge potential in the supergravity multiplet of new minimal supergravity and thus do provide a candidate that could be equivalent to \eqref{a1}.
The candidate anomalies of type I, IVa, Va and Vb each involve at least two different Yang-Mills gauge fields and thus neither provide a candidate that could be equivalent to \eqref{a1}. 

Therefore, the only candidate anomalies derived in \cite{Brandt:1996au} that could possibly provide a candidate \eqref{a1} are those of type II. The integrands of type II candidates are of the form
\bea
(C^{i_a}\5\cD^2+\ldots)(\cA(\5 W,\5\lambda)+\cD^2\cB(\7\cT))+c.c.
\label{a2}
\eea
where $\cD^2=\cD^\alpha\cD_\alpha$ and $\5\cD^2=\5\cD_\da\5\cD^\da$ are operators that are quadratic in the supersymmetry transformations $\cD_\alpha$ and $\5\cD_\da$ of supercovariant tensors, respectively, $C^{i_a}$ is an abelian Yang-Mills ghost,
$\cA(\5 W,\5\lambda)$ is a polynomial in the antichiral part $\5W_{\da\dbe\dga}$ of the supercovariant gravitino field strength (cf. eq. (3.1) of \cite{Brandt:1996au}) and in the gauginos $\5\lambda_\da{}^i$, and $\cB(\7\cT)$ is a polynomial in the supercovariant tensors $\7\cT^r$ given in equation (6.2) of \cite{Brandt:1996au} (chiral matter multiplets can and will be ignored for the discussion of potential candidate anomalies \eqref{a1}) where $\cA(\5 W,\5\lambda)$ and $\cB(\7\cT)$ are invariant under Lorentz transformations and gauge transformations of the Yang-Mills gauge group, except that $\cA(\5 W,\5\lambda)$ must have R-charge $-2$ if R-transformations are gauged. The full expression for the type II candidate anomalies can be found in equation (3.12) of \cite{Brandt:1993vd} but it will not be relevant here.

Now, it is easy to verify that \eqref{a2} does not provide a local function that could be equivalent to the integrand of \eqref{a1}, using the following ``natural'' mass dimensions:
\bea
&&[\viel a\mu]=0,\ [\RS \alpha\mu]=[\cD_\alpha]=1/2,\ [A_\mu{}^i]=[\6_\mu]=1,\ [\lambda_\alpha{}^i]=3/2,
\nonumber\\
&&[C^I]=0,\ [C^\mu]=[dx^\mu]=-1,\ [\xi^\alpha]=-1/2
\label{d1a}
\eea
where $\viel a\mu$ is the vierbein, $A_\mu{}^i$ are the Yang-Mills gauge fields,
$C^I$ are the Lorentz and Yang-Mills ghosts and $C^\mu$ are the ghosts of general coordinate transformations.
With these definitions the integrand of \eqref{a1} has mass dimension 3 (disregarding $[d^4x]$ here). Hence, in order that  \eqref{a2} has the same mass dimension 3 as the integrand of  \eqref{a1}, the polynomial $\cA(\5 W,\5\lambda)$ in \eqref{a2} would have to have mass dimension 2 and the polynomial $\cB(\7\cT)$ would have to have mass dimension 1. However, there are no polynomials $\cA(\5 W,\5\lambda)$ and $\cB(\7\cT)$ with these mass dimensions because 
$\5W_{\da\dbe\dga}$ and $\5\lambda_\da{}^i$ both have mass dimension 3/2 (recall that $\5W_{\da\dbe\dga}$ is the antichiral part of the supercovariant gravitino field strength and thus has the mass dimension  of $\6_\mu\bRS \da\nu$), and all tensors $\7\cT^r$ in equation (6.2) of \cite{Brandt:1996au} have mass dimensions $\geq 3/2$.\footnote{The reasoning analogously applies in the presence of chiral matter multiplets because \eqref{d1a} only fixes the mass dimensions of the component fields of chiral matter multiplets relatively to each other, but not relatively to other fields. Therefore, one can assign arbitrary mass dimension to the lowest component fields of chiral matter multiplets consistently with \eqref{d1a}.}

Assuming that the results of \cite{Brandt:1996au} concerning candidate anomalies are correct, we conclude that there is no candidate anomaly of the form of \eqref{a1}. In other words, any local functional \eqref{a1} whose integrand is BRST-invariant modulo a total divergence is trivial, i.e. its integrand is BRST-exact modulo a total divergence.

Of course, this reasoning relies on the correctness of the results of \cite{Brandt:1996au} concerning candidate anomalies. I am convinced that these results are correct. However, I understand that some reader may have doubts concerning these results and may not be willing to go through all the admittedly (but inevitably) rather complex derivations in \cite{Brandt:1996au}. Therefore I shall outline now that and how
one can conclude that there is no candidate anomaly of the form of \eqref{a1} without using the more sophisticated results of \cite{Brandt:1996au} by showing that one can derive this conclusion
already by techniques and results that are well-established in local BRST cohomology, namely the descent equation technique, contracting homotopies to eliminate ``trivial pairs'' (also called ``contractible pairs''), properties of the Koszul-Tate differential contained in the BRST-differential, and standard Lie algebra cohomology (cf. \cite{Barnich:2000zw,Dragon:2012au} for reviews), in combination with the cohomology of an operator $\delta_-$ outlined below. Again, for this reasoning the mass dimension of the integrand of \eqref{a1} is crucial because it implies that the more sophisticated results of \cite{Brandt:1996au} are not needed to conclude that there is no candidate anomaly of the form of \eqref{a1}.

In the following I sketch steps (a) to (d) used  in \cite{Brandt:1996au} that lead to this conclusion.
Of course, it cannot be the purpose of this note to review or repeat these steps in detail.

(a) A local functional whose integrand is BRST-invariant up to a total divergence gives rise by the so-called descent equations to a cocycle of $\4s=s+d$ in the space of local ``total forms''   where $s$ is the BRST-differential, $d=dx^\mu\6_\mu$ is the exterior derivative and a total form  $\omega=\sum_p  \omega_p$ is the sum of local exterior forms  $\omega_p$ of the fields, antifields and their derivatives with various degrees $p$ and ghost numbers (the forms $\omega_p$ of a cocycle of $\4s$ are the forms that occur in the descent equations). If the integrand of the local functional has ghost number $g$, the corresponding total form has total degree $g+4$ (= sum of the form degree and the ghost number of its various exterior forms $\omega_p$)  because the volume element $d^4x$ has total degree 4. Furthermore, the integrand of the local functional is BRST-exact up to a total divergence if and only if the corresponding cocycle of $\4s$ is $\4s$-exact. Hence, in four dimensions the candidate anomalies arise from the cohomology of $\4s$ at total degree 5.

(b) The cohomology of $\4s$ locally\footnote{There are some ``topological'' $\4s$-cocycles that are locally but not globally $\4s$-exact, cf. section 11 of \cite{Brandt:1996au}. However, these are not relevant to the question whether or not there is a candidate anomaly of the form of \eqref{a1}.} is equivalent to the weak cohomology (= cohomology on-shell) of $\4s$ in the space of total forms $\omega(\4C,\4\xi,\7\cT)$ which only depend on the supercovariant tensors $\7\cT^r$ and on ``generalized connections'' $\4C^i$, $\4C^{ab}$, $\4\xi^a$, $\4\xi^\alpha$ and $\4\xi^\da$ defined according to
\bea
&&\4C^i=C^i+(C^\mu+dx^\mu)A_\mu{}^i,\ \4C^{ab}=C^{ab}+(C^\mu+dx^\mu)\om_\mu{}^{ab},\nonumber\\
&&\4\xi^a=(C^\mu+dx^\mu)\viel a\mu,\ 
\4\xi^\alpha=\xi^\alpha+(C^\mu+dx^\mu)\RS \alpha\mu,\ 
\4\xi^\da=\5\xi^\da-(C^\mu+dx^\mu)\bRS \da\mu \quad
\label{d8}
\eea
where $C^{ab}$ are the Lorentz ghosts and $\om_\mu{}^{ab}$ is the spin connection constructed of the vierbein and the gravitino.\footnote{E.g., total forms 
 $\omega(\4C,\4\xi,\7\cT)$ that correspond to the first and second term in \eqref{a1} are
 $\Xi W_{\alpha\beta\gamma}\4\xi^\alpha G^{\beta\gamma}$ and
 $\Xi \4C W_{\alpha\beta\gamma}W^{\alpha\beta\gamma}$, respectively, 
where $\Xi =(1/24)\epsilon_{abcd}\4\xi^a\4\xi^b\4\xi^c\4\xi^d$, and $G^{\beta\gamma}$ and $\4C$ correspond to $\6_\rho A_\sigma-\6_\sigma A_\rho$ and $C$, respectively.}
 This can be shown by means of contracting homotopies which eliminate ``trivial pairs'' from the cohomology of $\4s$ and by means of the properties of the Koszul-Tate differential contained in $\4s$ which allow to remove  from the antifield independent part of a cocylce of $\4s$ all supercovariant tensors (such as the supercovariantized Ricci tensor or the auxiliary fields) which vanish on-shell or are equal on-shell to polynomials in the $\7\cT$'s, cf., e.g., \cite{Brandt:1996mh}.
 
(c) The results  of standard Lie algebra cohomology imply then that any nontrivial cocycle $\omega(\4C,\4\xi,\7\cT)$ of the weak cohomology of $\4s$ can be assumed to contain a term\footnote{We are using old minimal supergravity here which is compatible with \cite{Dixon:2021ydu}. Therefore $\4Q$ which may contribute in eq. (6.5) of \cite{Brandt:1996au} in the case of new minimal supergravity is not present in \eqref{d9}.}
\bea
f_i(\4\xi,{\7\cT})P^i(\4\theta)
\label{d9}
\eea
where the $\4\theta$'s are polynomials in the $\4C$'s of the Lie algebra cohomology (cf. eq. (4.23) of \cite{Brandt:1996au}) and the
$f_i(\4\xi,{\cal T})$ are polynomials in the $\4\xi$'s and $\7\cT$'s that are nontrivial cocycles of the weak cohomology of $\4s$ in the space of total forms $f(\4\xi,\7\cT)$:
\bea
\4s f_i(\4\xi,{\7\cT})\approx 0,\qd f_i(\4\xi,{\7\cT})\not\approx \4s g_i(\4\xi,{\7\cT}).
\label{d10}
\eea
The first eq. \eqref{d10} particularly imposes invariance of $f_i(\4\xi,{\7\cT})$ 
under Lorentz and Yang-Mills gauge transformations because $\4sf_i(\4\xi,{\7\cT})$ contains $\4C^I\delta_If_i(\4\xi,{\7\cT})$.

(d) The problem imposed by eqs. \eqref{d10} is the most involved part of the computation of the local BRST-cohomology in supergravity, and the general solution of this problem is the key result of the computation. It is this problem where the structure of the supersymmetry multiplets formed by the $\7\cT$'s (termed ``QDS-structure'' in \cite{Brandt:1996au}) comes into play. However, as will be shown now, this structure is not needed in order to conclude the absence of candidate anomalies of the form of \eqref{a1}. The strategy used in \cite{Brandt:1996au} to solve the problem imposed by eqs. \eqref{d10} is a decomposition according to the degree in the $\4\xi^a$. $\4s$ has precisely one part which decrements this degree. This part is the operator 
$\delta_-$ given by
\bea
\delta_-=2i\4\xi^\alpha\sigma^a{}_{\alpha\da}\4\xi^\da\frac{\6}{\6\4\xi^a}\ .
\label{d11}
\eea
This implies that the part of lowest degree in the $\4\xi^a$ contained in a solution $f_i$ of eqs. \eqref{d10} can be assumed to be a nontrivial cocycle of the cohomology of $\delta_-$. This cohomology is summarized in eqs. (E.13) and (E.14) of \cite{Brandt:1996au}. Notice that $\delta_-$ involves only the $\4\xi$'s. Therefore, the supersymmetry multiplet structure of the 
$\7\cT$'s does not matter at all to the cohomology of $\delta_-$. By this cohomology the part of lowest degree in the $\4\xi^a$ contained in a solution $f_i$ of eqs. \eqref{d10} can be assumed to be of the form 
\bea
P(\5\vartheta^\da,\4\xi^\alpha,\7\cT)+P^\prime(\vartheta^\alpha,\4\xi^\da,\7\cT)+\Theta R(\7\cT)
\label{d3}
\eea
where
\bea
\vartheta^\alpha=\4\xi_\da\4\xi^{\alpha\da},\ 
\5\vartheta^\da=\4\xi^{\alpha\da}\4\xi_\alpha,\ 
\Theta=\4\xi_\da\4\xi^{\alpha\da}\4\xi_\alpha
\label{d4}
\eea
and $P(\5\vartheta^\da,\4\xi^\alpha,\7\cT)$, $P^\prime(\vartheta^\alpha,\4\xi^\da,\7\cT)$ and $R(\7\cT)$ are polynomials in their arguments. 

The non-existence of a candidate anomaly of the form of \eqref{a1} is implied by the fact that a polynomial \eqref{d3} which might correspond to such a candidate anomaly simply does not exist. Again, the reason is the mass dimension of the integrand of \eqref{a1}. In this context, the relevant mass dimension is $-1$ because here we are discussing the descent equations which involve the differentials $dx^\mu$ and therefore the mass dimension $[d^4x]=-4$ of the volume element of the integrand of \eqref{a1} has to be taken into account. Furthermore, as noted already above, a total form that corresponds to a candidate anomaly has total degree 5 (which is the total degree of the integrand of \eqref{a1}, taking into account the volume element $d^4x$). Therefore any nontrivial $\4s$-cocycle that might correspond to a candidate anomaly of the form of \eqref{a1} can be assumed to contain a polynomial \eqref{d3} with mass dimension $-1$ and total degree $\leq 5$ because the $P^i(\4\theta)$ in \eqref{d9} have vanishing mass dimension and total degrees $\geq 0$ (a $P^i$ with total degree 0 is a constant). It is easy to verify that such polynomials \eqref{d3}  do not exist, using that the only relevant $\7\cT$'s (those with sufficiently low mass dimension) are:
\bea
\ba{c|c}
\7\cT & [\7\cT] \\
\hline\rule{0em}{3ex}
\lambda_\alpha{}^i , \5\lambda_\da{}^i , W_{\alpha\beta\gamma}, \5W_{\da\dbe\dga}& 3/2 \\
G_{\alpha\beta}{}^i ,\5G_{\da\dbe}{}^i, X_{\alpha\beta\gamma\delta}, \5X_{\da\dbe\dga{\dot \delta}}& 2 \\
\cD_{(\alpha}{}^\da\lambda_{\beta)}{}^i , \cD^\alpha{}_{(\da}\5\lambda_{\dbe)}{}^i,
\cD_{(\alpha}{}^\da W_{\beta\gamma\delta)}, \cD^\alpha{}_{(\da} \5W_{\dbe\dga{\dot \delta})} & 5/2
\ea
\label{d6}
\eea
where $G_{\alpha\beta}{}^i$ and $X_{\alpha\beta\gamma\delta}$ are the components with undotted spinor indices of the supercovariantized Yang-Mills field strengths and Weyl tensor, respectively (cf. eqs. (3.2) and (3.3) of \cite{Brandt:1996au}), and $\cD_{\alpha\da}$ are the supercovariant derivatives written with spinor indices.
E.g. the most general Lorentz-invariant polynomial $P(\5\vartheta^\da,\4\xi^\alpha,\7\cT)$ in (\ref{d3}) with total degree 5 and mass dimension $-1$ would be of the form (notice that the $\5\vartheta^\da$ anticommute, i.e. there can be at most two of them and if there are two of them they give a term $\5\vartheta_\da\5\vartheta^\da$)
\bea
\5\vartheta_\da\5\vartheta^\da \4\xi^\alpha f_\alpha^{5/2}(\7\cT)+
\5\vartheta^\da\4\xi^\alpha\4\xi^\beta\4\xi^\gamma f_{\alpha\beta\gamma\da}^{2}(\7\cT)+
\4\xi^\alpha\4\xi^\beta\4\xi^\gamma\4\xi^\delta\4\xi^\epsilon f_{\alpha\beta\gamma\delta\epsilon}^{3/2}(\7\cT)
\label{d5}
\eea
where the superscript of an $f$ indicates its mass dimension and the $f$'s are totally symmetric in all their undotted spinor indices (because the $\4\xi^\alpha$ commute). 
It can be easily checked that with the tensors \eqref{d6} it is impossible to construct polynomials $f_\alpha^{5/2}(\7\cT)$, 
$f_{\alpha\beta\gamma\da}^{2}(\7\cT)$ and $f_{\alpha\beta\gamma\delta\epsilon}^{3/2}(\7\cT)$ with the mass dimensions indicated by the superscripts and the Lorentz structure indicated by the spinor indices. 
 In the same way one finds that no Lorentz-invariant polynomials (\ref{d3}) with mass dimension $-1$ and total degrees 0 to 4 exist that can be constructed of the tensors (\ref{d6}).


\begin{thebibliography}{99}

\bibitem{Dixon:2021ydu}
J.~A.~Dixon,
``A candidate for a supergravity anomaly,''
[arXiv:2110.06213v1 [physics.gen-ph]]

\bibitem{Brandt:1996au}
F.~Brandt, 
``Local BRST cohomology in minimal D=4, N=1 supergravity,''
  Annals Phys. 259 (1997) 253
[arXiv:hep-th/9609192]

\bibitem{Stelle:1978ye}
K.~S. Stelle and P.~C. West, 
``Minimal auxiliary fields for supergravity,''
  Phys. Lett. B 74 (1978) 330

\bibitem{Ferrara:1978em}
S.~Ferrara and P.~van Nieuwenhuizen, 
``The auxiliary fields of supergravity,''
Phys. Lett. B 74 (1978) 333

\bibitem{Sohnius:1981tp}
M.~F. Sohnius and P.~C. West, 
``An alternative minimal off-shell version of
  N=1 supergravity,'' 
  Phys. Lett. B 105 (1981) 353

\bibitem{Brandt:1993vd}
F.~Brandt, 
``Anomaly candidates and invariants of D=4, N=1 supergravity theories,''
Class. Quant. Grav. 11 (1994) 849
[arXiv:hep-th/9306054]

\bibitem{Barnich:2000zw}
G.~Barnich, F.~Brandt and M.~Henneaux,
``Local BRST cohomology in gauge theories,''
Phys. Rept. \textbf{338} (2000) 439
[arXiv:hep-th/0002245]

\bibitem{Dragon:2012au}
N.~Dragon and F.~Brandt,
``BRST symmetry and cohomology,''
in: Strings, Gauge Fields, and the Geometry Behind, L. Katzarkov, J. Knapp and A. Rebhan (eds.), World Scientific (2012)
[arXiv:1205.3293 [hep-th]]

\bibitem{Brandt:1996mh}
F.~Brandt,
``Local BRST cohomology and covariance,''
Commun. Math. Phys. 190 (1997) 459
[arXiv:hep-th/9604025]

\end{thebibliography}
\end{document}